\documentclass[final,3p,times,twocolumn]{elsarticle}

\usepackage{amssymb}
\usepackage{braket}
\usepackage{amsmath}
\usepackage{xspace}
\usepackage{color}

\newcounter{bla}

\newcommand{\TeNeS}{\textsf{TeNeS}\xspace}
\newcommand{\tenes}{\texttt{tenes}\xspace}
\newcommand{\tenesstd}{\texttt{tenes\_std}\xspace}
\newcommand{\tenessimple}{\texttt{tenes\_simple}\xspace}

\journal{Computer Physics Communications}

\begin{document}

\begin{frontmatter}



\title{\TeNeS-v2: Enhancement for Real-Time and Finite Temperature Simulations of Quantum Many-Body Systems}


\author[a]{Yuichi Motoyama\corref{author}}
\author[b]{Tsuyoshi Okubo\corref{author}}
\author[a]{Kazuyoshi Yoshimi}
\author[c]{Satoshi Morita}
\author[a]{Tatsumi Aoyama}
\author[a]{Takeo Kato}
\author[a]{Naoki Kawashima}

\cortext[author] {Corresponding author.\\\textit{E-mail address:} y-motoyama@issp.u-tokyo.ac.jp (Yuichi Motoyama), t-okubo@phys.s.u-tokyo.ac.jp (Tsuyoshi Okubo)}
\address[a]{The Institute for Solid State Physics, The University of Tokyo, Chiba 277-8581, Japan}
\address[b]{Institute for Physics of Intelligence, The University of Tokyo, Tokyo 113-0033, Japan}
\address[c]{Graduate School of Science and Technology, Keio University, Kanagawa 223-8522, Japan}

\begin{abstract}
Quantum many-body systems are challenging targets for computational physics due to their large degrees of freedom. The tensor networks, particularly Tensor Product States (TPS) and Projected Entangled Pair States (PEPS), effectively represent these systems on two-dimensional lattices. 
However, the technical complexity of TPS/PEPS-based coding is often too much for researchers to handle. To reduce this problem, 
we developed \TeNeS (Tensor Network Solver). This paper introduces \TeNeS-v2, which extends \TeNeS with real-time and finite temperature simulations, providing deeper insights into quantum many-body systems. We detail the new algorithms, input/output design, and application examples, demonstrating \TeNeS-v2's applicability to various quantum spin and Bose models on two-dimensional lattices.
\end{abstract}


\end{frontmatter}



{\bf NEW VERSION PROGRAM SUMMARY}

\begin{small}
\noindent
{\em Program Title:} \TeNeS                                        \\
{\em CPC Library link to program files:} (to be added by Technical Editor) \\
{\em Developer's repository link:} https://github.com/issp-center-dev/TeNeS \\
{\em Code Ocean capsule:} (to be added by Technical Editor)\\
{\em Licensing provisions:} GPLv3  \\
{\em Programming language:} C++11                             \\
{\em Journal reference of previous version:} doi:10.1016/j.cpc.2022.108437 \\
{\em Reasons for the new version:} To extend the capabilities of \TeNeS to include real-time and finite temperature simulations, enabling deeper insights into quantum many-body systems beyond ground states. \\
{\em Summary of revisions:} \TeNeS-v2 introduces real-time evolution and finite temperature simulation capabilities, expanding the range of quantum many-body phenomena that can be studied.\\
{\em Nature of problem(approx. 50-250 words):} Quantum many-body systems are extremely difficult to simulate because of the huge dimension of the Hilbert space. Conventional methods have difficulty in accurately representing these systems, especially for properties beyond small lattices and ground states. Advanced computational techniques are required to efficiently capture the dynamics and thermal properties of these systems. \\
{\em Solution method(approx. 50-250 words):}\TeNeS employs tensor networks, specifically Tensor Product States (TPS) and Projected Entangled Pair States (PEPS), to efficiently represent quantum many-body states on two-dimensional lattices. The real-time evolution is handled through a straightforward extension of imaginary time evolution methods, allowing the study of dynamical properties. Finite temperature simulations are conducted using an imaginary time evolution approach starting from the infinite-temperature mixed state. These methods enable the accurate calculation of various physical properties in different quantum states.\\
{\em Additional comments including restrictions and unusual features (approx. 50-250 words):}
\TeNeS-v2 maintains ease of use by providing flexible input file configurations and supporting a variety of two-dimensional lattice models. Increasing the bond dimension to improve accuracy requires more computational resources. Finite temperature calculations may exhibit unphysical behavior due to limitations of the tensor network approximation method. Users should be aware of these potential problems and interpret the results following the suggestions made in the text.\\


\end{small}


\section{Introduction}
\label{sec:intro}
Quantum many-body systems are challenging targets for computational physics due to their huge degrees of freedom. Typically, Hilbert space dimensions exponentially increase with the number of particles, and solving exactly an eigenvalue problem for a given Hamiltonian becomes impossible for large systems. The tensor network approach is one of the promising tools to tackle such quantum many-body problems through data compression based on tensor networks \cite{Orus2014,Orus2019}. In the tensor network approach, we represent a quantum many-body state by a network of small tensors connected by contractions of tensor legs. Thanks to the area law of the quantum entanglement \cite{EisertCP2010}, such tensor network states (TNS) are efficient representations for low energy states, and they have been widely used as a variational wave function, typically for the ground states. 

For models defined on a two-dimensional lattice, the Tensor Product State (TPS) \cite{Nishino2001} (also known as Projected Entangled Pair State (PEPS) \cite{Verstraete2004})) has been the most successful tensor network representation. It has been successfully applied to ground states of a variety of spin models \cite{Corboz2013, Okubo2017, Liao2017, LeeKOK2019, Lee2020a, MashikoO2024} and fermion systems \cite{Corboz2014, Corboz2014_2}. Recently, TPS/PEPS have also been extended to real-time simulations\cite{Czarnik2019, HubigBKGC2020,KanekoD2022, PonnagantiMP2023}, excited states \cite{Ponsioen2020,TuVSLKC2024}, and finite temperatures \cite{Czarnik2019, Kshetrimayum2019, CzarnikRCD2021,Jimenez2021}. From simulations of real-time dynamics and finite temperature properties, one can extract information of excitations in quantum many-body systems. Furthermore, they are suitable for comparison with the experimental data. However, despite the success of TPS/PEPS simulations, it is still complicated to develop the TPS/PEPS-based programs due to many tensor network contractions and tensor optimizations in the algorithms. 

To overcome the difficulties in using TPS/PEPS simulations, we have developed \TeNeS (Tensor Network Solver) designed to allow easy access to tensor network calculations, based on TPS/PEPS \cite{TeNeS_Github,TeNeS_webpage,TeNeSv1}. As we describe in detail in the following sections, \TeNeS is an application for investigating a variety of quantum spin and boson models on arbitrary two-dimensional lattices, based on the infinite version of TPS/PEPS called iTPS/iPEPS \cite{Orus2009}. In our previous paper \cite{TeNeSv1}, we introduced the calculation of the ground state based on the imaginary time evolution. In this paper, we explain the new features in \TeNeS-v2: real-time and finite temperature simulations, which enable us to extract richer information about quantum many-body systems beyond the ground state.

The remainder of the paper is organized as follows. We describe the functions of the previous version of \TeNeS in Section~\ref{sec:about}. In Section~\ref{sec:newfeatures}, we explain the new features and their algorithms in \TeNeS-v2. Section~\ref{sec:input} is devoted to the design of input/output for the new features. Examples of applications are presented in Section~\ref{sec:examples}. Finally, we conclude our work in Section~\ref{sec:conclusion}.  

\section{About \TeNeS-v1} 
\label{sec:about}

Before explaining the new features of the latest version of \TeNeS, we briefly summarize the basic functions that have already been implemented in the previous version~\cite{TeNeS_Github, TeNeS_webpage, TeNeSv1}.
As a basic functionality, \TeNeS performs a calculation of the ground state of two-dimensional quantum lattice systems in the thermodynamic limit.
In \TeNeS, the wave function of the spin or boson systems is expressed with the infinite version of the tensor product states (iTPS) (see Fig.~\ref{fig:iTPO}(a)) and is optimized by the imaginary time evolution method.
For contraction of the tensors, the corner transfer matrix renormalization group (CTMRG) method and the mean-field environment (MFE) method are supported and users can select one of these.
Although \TeNeS can treat iTPS only on a square lattice, various lattice models can be treated by supporting interactions beyond the nearest neighbors.

The procedure for optimizing the tensors for the ground state of a given Hamiltonian $H$ is summarized as follows.
\TeNeS prepares an appropriate initial wave function $|\Psi_0\rangle$ and
uses the imaginary time evolution,
\begin{equation}
    |\Psi^{\mathrm{iTPS}}\rangle \simeq e^{-T\mathcal{H}}|\Psi_0\rangle,
\end{equation}
for a sufficiently long imaginary time $T$.
The Hamiltonian $\mathcal{H}$ is assumed to be composed of short-range two-body interactions, $H_{ij}$.
Using the Suzuki--Trotter decomposition, the imaginary time evolution is represented as
\begin{equation}
    e^{-T\mathcal{H}} |\Psi_0\rangle= \left(\prod_{\{(i,j)\}} e^{-\tau H_{ij}}\right)^{N_\tau} |\Psi_0\rangle+ O(\tau),
\end{equation}
where $N_\tau$ is the division number of time and $\tau = T/N_\tau$ is a sufficiently small time interval.
In multiplying the operator $e^{-\tau H_{ij}}$, the wave function is truncated to a finite bond dimension $D$ in the iTPS.
We call multiplication of the exponential function of the Hamiltonian and subsequent truncation as an {\it update}.
\TeNeS implements two types of update methods, i.e., the standard simple update and full update methods for truncations~\cite{Jiang2008, Jordan2008, Orus2009, Phien2015}.
In both methods, the bond dimension is truncated so as to minimize the norm of difference,
\begin{equation}
\| \Ket{\Psi^\text{iTPS}_\text{new}} - e^{-\tau \mathcal{H}_{ij}}\Ket{\Psi^\text{iTPS}} \|.
\end{equation}
\TeNeS first performs simple updates and then full updates in the optimization process.
The number of these updates can be controlled by the users.

The main program of \TeNeS (\tenes) optimizes an iTPS based on the imaginary time method and evaluates the expectation values of physical quantities. 
By preparing the input file, the user can flexibly define observables and various two-dimensional lattice models. 
For ease of use, the user can also use two additional scripts, \tenesstd and \tenessimple, to automatically generate the input files.
By using these programs, users can easily calculate the expectation values of local quantities described by the spin (boson) operators for the ground state based on the iTPS method.
\TeNeS includes several samples of the input files for calculation of the transverse-field ferromagnetic Ising model, $S=1/2$ Heisenberg model, and hardcore boson model. 

\section{New features}
\label{sec:newfeatures}

\subsection{Time evolution}
\label{sec:time_evolv}

Under a time-independent Hamiltonian $\mathcal{H}$, a wave function evolves from an initial state $\ket{\psi(t=0)}$ as
\begin{equation}
\Ket{\psi(t)} = e^{-i\mathcal{H}t/\hbar}\Ket{\psi(0)}.
\end{equation}
As for the ground state calculation, the wave function can be represented by the iTPS, and the time evolution is achieved by the Suzuki-Trotter decomposition and the simple/full update methods.
Technically, the real-time evolution can be calculated in the same way as the ground state calculation simply by changing the coefficient of the time evolution operator; $\tau \to -it/\hbar$.

For the real-time evolution calculation mode, \tenesstd calculates the real-time evolution operator instead of the imaginary time evolution operator (note that $\hbar=1$ is used).
Additionally, \tenes calculates observables along time evolution as
\begin{equation}
\Braket{A(t)} = \frac{\braket{\psi(t)|\hat{A}|\psi(t)}}{\braket{\psi(t)|\psi(t)}}.
\end{equation}
Note that the required bond dimensions increase in general as the quantum entanglement grows over time.

\subsection{Finite temperature}
\label{sec:finite_temp}

\begin{figure}[t]
  \begin{center}
    \includegraphics[width=1.0\columnwidth]{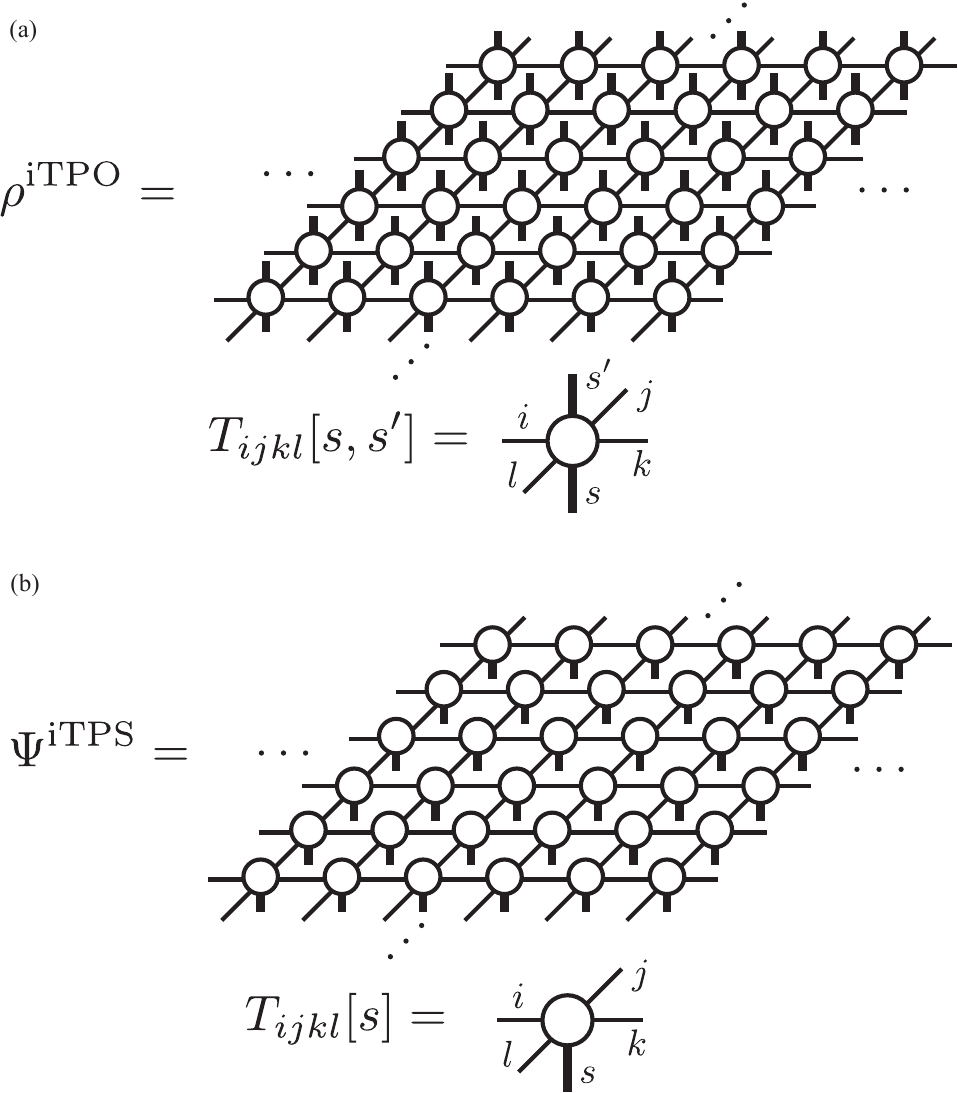}
    \caption{(a)Tensor network diagram of iTPO for the model on the infinite square lattice. (b) Similar diagram of iTPS for a pure state.}
    \label{fig:iTPO}
  \end{center}
\end{figure}

At a finite temperature, the quantum state is usually represented by a mixed state instead of a pure state. For a given Hamiltonian $\mathcal{H}$ and an inverse temperature $\beta = 1/T$, the density operator representing a mixed state at temperature $T$ is given by
\begin{equation}
   \rho(\beta) = \frac{e^{-\beta \mathcal{H}}}{\mathrm{Tr} e^{-\beta \mathcal{H}}}.
\end{equation}
We can consider a tensor network representation for a mixed state similar to the case of pure states. As an example, let us consider a quantum spin system with $S=1/2$ at a finite temperature. For a $N$-spin system, the mixed state is expressed as 
\begin{equation}
       \rho(\beta) = \sum_{s_i=\uparrow, \downarrow, s_i' = \uparrow, \downarrow} \left(\rho(\beta)\right)_{s_1,s_2,\dots, s_N}^{s_1', s_2', \dots, s_N'} |s_1', s_2', \dots, s_N'\rangle \langle s_1, s_2, \dots, s_N|,
       \label{eq:rho}
\end{equation}
where the coefficient $\left(\rho(\beta)\right)_{s_1,s_2,\dots, s_N}^{s_1', s_2', \dots, s_N'}$ in the right hand side is the density matrix. We can apply a tensor network approximation for this density matrix. The tensor network diagram of an infinite Tensor Product Operator (iTPO) representing a mixed state for the model on the infinite square lattice is given in Fig.~\ref{fig:iTPO}(a). The two bold lines extending from each circle represent local spin indices, $s_i$ and $s_i'$ in Eq.~\eqref{eq:rho}. The other four thin lines indicate virtual indices. Similar to iTPS for the approximation of the ground state, the accuracy of iTPO is controlled by the bond dimension $D$, which is the dimension of the virtual indices ($i$, $j$, $k$, $l$) in Fig.~\ref{fig:iTPO}(a).

We can calculate the density matrix at an inverse temperature $\beta = 1/T$ by the imaginary time evolution starting from $\beta = 0$. Note that the initial density matrix is the identity matrix and can be represented by $D=1$ iTPO, which is the tensor product of local identity matrices $\delta_{s_i, s_i'}$. The imaginary time evolution of iTPO is calculated by a simple extension of the imaginary time evolution for iTPS. The Suzuki-Trotter decomposition, the simple update method, and the full update method used for pure states can be applied almost directly to the case of iTPO. \TeNeS currently supports only the simple update because the full update for iTPO consumes huge computational resources. 

\begin{figure}[t]
  \begin{center}
    \includegraphics[width=1.0\columnwidth]{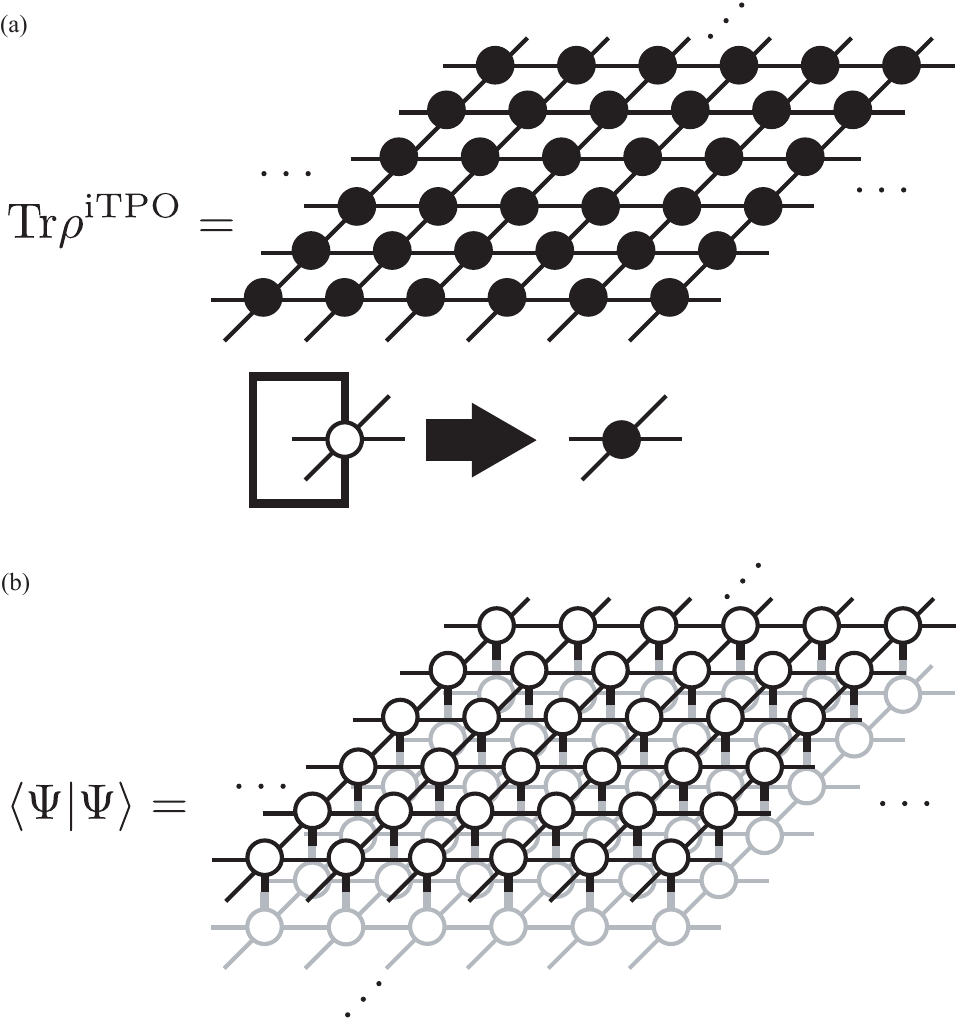}
    \caption{(a) Single layer structure of the trace of iTPO. (b) Double layer structure for the inner product of iTPS.}
    \label{fig:iTPO_trace}
  \end{center}
\end{figure}

\begin{figure}[t]
  \begin{center}
    \includegraphics[width=1.0\columnwidth]{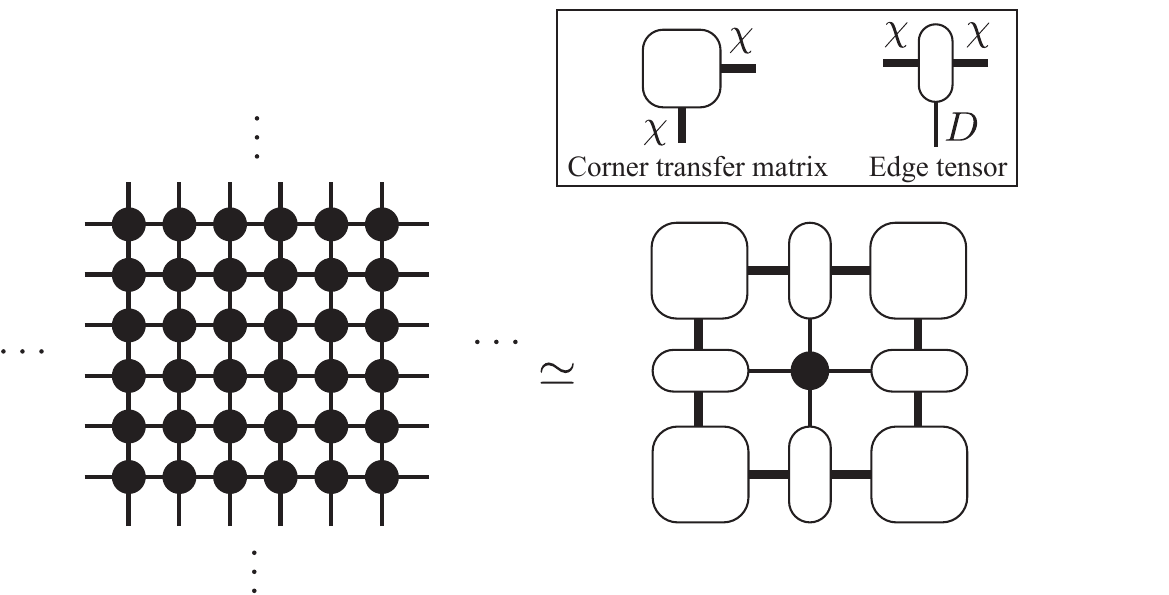}
    \caption{Corner transfer matrix representation of the single layer network for the expectation value calculation with iTPO. The quality of the approximation is controlled by the bond dimension $\chi$ of the corner transfer matrices.}
    \label{fig:CTMRG}
  \end{center}
\end{figure}

The expectation value of a physical quantity $O$ for a given mixed state described by $\rho$ is given as
\begin{equation}
   \langle O \rangle_\rho = \frac{\mathrm{Tr}\, (\rho O)}{\mathrm{Tr}\, \rho}.
\end{equation}
The trace $\mathrm{Tr}$ corresponds to connecting the corresponding upper and lower legs of the iTPO, namely, physical indices $s_i$ and $s_i'$. Using a tensor obtained by contracting the physical indices in iTPO, the denominator $\mathrm{Tr} \, \rho$ becomes the two-dimensional square lattice 
tensor network (See Fig.~\ref{fig:iTPO_trace}(a)), which is identical with the one that appeared in the expectation values for pure states. Thus, we can apply the same approximate calculation using corner transfer matrix representation and CTMRG (See Fig.~\ref{fig:CTMRG}). Similarly to the case of pure states, once the converged corner transfer matrices and edge tensors are computed, $\mathrm{Tr} \, (\rho O)$ can also be efficiently calculated. 
 
In the case of finite temperature calculation with bond dimension $D$ iTPO, the computational cost of CTMRG for the corner transfer matrix representation with bond dimension $\chi$ scales with the larger of $O(\chi^2 D^4)$ and $O(\chi^3 D^3)$.
Empirically, the method performs best when $\chi$ is taken to be proportional to $D$, \textit{i.e.}, $\chi \propto O(D)$. Under this condition, the computation cost of CTMRG becomes $O(D^6)$, and the required memory amount becomes $O(D^4)$. 

One significant difference between the computations of finite temperature states by iTPO and pure states by iTPS is in the tensor network contraction for expectation values. In the case of iTPO, the square lattice tensor network for the expectation value calculation consists of a single layer of bond-dimension $D$ tensors (Fig.~\ref{fig:iTPO_trace}(a)), while it is represented by a double layer in the case of iTPS (Fig.~\ref{fig:iTPO_trace}(b)). Due to this difference, the computational cost of finite temperature calculations using iTPO is significantly lower than that of the ground state calculation using iTPS with the same bond dimension $D$. It allows us to use larger bond dimensions in finite temperature calculations. Note that in the limit of zero temperature, iTPO is expected to become a pure state corresponding to the bond dimension $\sqrt{D}$ iTPS. Therefore, there is no apparent benefit in using iTPO for ground state calculations. 

Lastly, we mention the drawbacks of the approximation with iTPO. The density matrix of a mixed state is Hermitian and positive semidefinite, with non-negative eigenvalues. However, when approximating the density matrix with iTPO, this positive semidefiniteness is not guaranteed, and physical quantities calculated from the iTPO approximation might exhibit unphysical behavior, such as energies lower than the ground state energy. This is a problem of iTPO representation and cannot be avoided just by improving the accuracy of CTMRG in the calculation of the expectation value by increasing the bond dimension $\chi$. To recover physical behavior, it is necessary to increase the bond dimension $D$ of iTPO to improve the accuracy of the density matrix.

As an alternative representation to avoid such unphysical behavior, a method using a purified representation of the density matrix with iTPO has been proposed \cite{Czarnik2019}. However, in this method, the diagram appearing in the expectation value calculation becomes a double-layer structure similar to that of pure states. This structure requires a larger computational cost, and the manageable bond dimension $D$ becomes smaller than in the single-layer iTPO representation that has been discussed above and is supported in TeNeS-v2.

\section{Input/Output files}
\label{sec:input}
\subsection{Real-time evolution}

A new parameter \texttt{parameter.general.mode} is introduced to control the kind of calculation that is performed in \TeNeS.
Its default value is \texttt{"ground state"}, which means the ground state calculation.
When \texttt{"time evolution"} (or \texttt{"time"}) is specified, \TeNeS simulates real-time evolution.
As for the ground state calculation, \texttt{tenes\_std} command calculates real-time evolutionary operators, $U_{ij}(t) = \exp[-it\mathcal{H}_{ij}]$, where the time step $t$ is specified by \texttt{parameter.simple\_update.tau} and \texttt{parameter.full\_update.tau}.
To change the time step during the simulation,
multiple values can be passed to time step \texttt{tau} and the number of steps \texttt{num\_step} like
\begin{verbatim}
num_step = [50, 200, 10]
tau = [0.01, 0.005, 0.05]
\end{verbatim}
This example means that the real-time evolution with $t=0.01$ is performed in the first 50 steps, the evolution with $t=0.005$ is done in the following 200 steps, and the evolution with $t=0.05$ is performed in the final 10 steps.
Another new parameter \texttt{parameter.general.measure\_interval} is the number of steps between the measurements of observables $n_\text{interval}$; \texttt{tenes} calculates the expectation values with every $n_\text{interval}$ updates.
As for \texttt{tau} and \texttt{num\_step}, multiple values can be set to \texttt{measure\_interval}.

The output files for the real-time evolution mode are named with the prefix \texttt{"TE\_"}.
Except for \texttt{TE\_density.dat}, the only difference from the ground state mode is that the time $t$ is inserted as the first column.
For example, \texttt{TE\_onesite\_obs.dat} that includes one-site observables such as $S_i^z$ has five columns like\footnote{We here reduce the number of decimal places for simplicity. In the actual output file, more decimal places are printed.}
\begin{verbatim}
0.5 0 1 4.63e-01 0.0
\end{verbatim}
The first column (0.5) is time, the second (0) is the observable ID, the third (1) is the index of the site, and the fourth (4.63e-01) and fifth (0.0) are the real and imaginary parts of the expectation value, respectively.
So the above example means $\braket{S^z}=0.463$ at $t=0.5$ on the site $1$.
The format of \texttt{TE\_density.dat} differs from that of \texttt{density.dat}.
While in \texttt{density.dat}, the densities of observables (observables par site) are saved in a key-value format like \texttt{Energy = 0.5 0.0} (the first and second figures correspond to the real and imaginary parts, respectively),
in \texttt{TE\_density.dat}, the densities are saved in the similar format of \texttt{TE\_onsite\_obs.dat} except that the spatial information such as the site index is dropped.

\subsection{Finite temperature calculation}

When \texttt{parameter.general.mode} is set to \texttt{"finite temperature"} (or \texttt{"finite"}) \texttt{tenes} performs the finite temperature calculation.
The other parameters such as \texttt{parameter.general.measure\_interval} are the same as for the real-time evolution mode.

The format of the output files is almost the same as that for the real-time evolution mode.
The differences are as follows; the prefix of the filenames is \texttt{"FT\_"} instead of \texttt{"TE\_"},
and the first column is the inverse temperature $\beta$ instead of the time $t$.
Because imaginary time evolutionary operators are applied for both the bra and the ket as
\begin{equation}
    \rho(2\tau) = e^{-\tau \mathcal{H}} \rho(0) e^{-\tau \mathcal{H}},
\end{equation}
it should be noted that the step size $\tau$ in the imaginary-time evolution operators corresponds to $2\tau$ in the actual inverse temperature step.

\subsection{Multi-sites observables}
\label{sec:multi_site}

In \TeNeS v1, only two-site observables such as spin-spin correlation $S_i^z S_j^z$ can be evaluated.
While it may be sufficient in many cases, observables that depend on three or more sites are desired in some cases, e.g., the scalar chirality for frustrated spin systems.
\TeNeS v2 can evaluate multi-site observables in the form of the summation of the product of one-site operators,
\begin{equation}
    O_{ij \dots k} = \sum_{a=1}^n \alpha^{(a)} A^{(a)}_i B^{(a)}_j \cdots C^{(a)}_k,
\label{eq:multisite}
\end{equation}
where $i, j, \dots, k$ are the indices of the sites,
$a$ is the index of the term,
$\alpha^{(a)}$ is the coefficient,
and $A^{(a)}, B^{(a)}, \dots C^{(a)}$ are the one-site operators like $S^z_i$.
For example, a scalar chirality of three spins $\chi_{ijk}$ is 
\begin{equation}
\begin{split}
\chi_{ijk} = \vec{S}_i \cdot \left(\vec{S}_j \times \vec{S}_k\right)
&=
S_i^x S_j^y S_k^z - S_i^x S_j^z S_k^y \\
&+
S_i^y S_j^z S_k^x - S_i^y S_j^x S_k^z \\
&+
S_i^z S_j^x S_k^y - S_i^z S_j^y S_k^x.
\end{split}
\label{eq:chirality}
\end{equation}
The following input implements this observable (for simplicity, we show the first two terms, $S_i^x S_j^y S_k^z$ and  $-S_i^x S_j^z S_k^y$):

\begin{verbatim}
[[observable.multisite]]
name = "Scalar"
group = 0
dim = [2, 2, 2]
ops = [1, 2, 0]
coeff = 1.0
multisites = """
0 1 0 0 1
"""

[[observable.multisite]]
name = "Scalar"
group = 0
dim = [2, 2, 2]
ops = [1, 0, 2]
coeff = -1.0
multisites = """
0 1 0 0 1
"""
\end{verbatim}
Here, \texttt{multisites} means the indices of the sites ($i, j, k$).
The first integer, 
\texttt{0} in the above example, means the first site index $i$, and the subsequent two integers are the relative coordinates ($dx=1$ and $dy=0$) of the second site $j$ measured from the first site.
The last two integers are the relative coordinates ($dx=0$, $dy=1$) of the third site $k$.
\texttt{ops} specifies the set of site operators defined in the \texttt{observable.onesite} section with the group ID.
\texttt{coeff} specifies the coefficient $\alpha^{(a)}$ in Eq.~(\ref{eq:multisite}).

The expectation value of the observable, $\braket{O_{ij\dots k}}$, is saved to the file \texttt{multisite\_obs\_N.dat}, where
\texttt{N} is the number of sites within the unit cell, for example, 3 for the scalar chirality.
The format of the file is similar to that of the two-site observable file, \texttt{twosite\_obs.dat}.

\section{Application examples}
\label{sec:examples}

Data and scripts for the following examples are available from the ISSP Data repository~\cite{ISSP-datarepo}.

\subsection{Real-time evolution}
\label{sec:time_evolv_example}

In the following, we present a computational demonstration that illustrates the real-time evolution of the Ising model on a square lattice under the influence of a transverse magnetic field $h_x$.
The Hamiltonian is given as follows:
\begin{equation}
\mathcal{H} = J^z \sum_{\langle i,j\rangle} S_i^z S_j^z - h(t)\sum_{i} S_{i}^{x},
\end{equation}
where $J^z = -1.0$ and $h(t) = h_x \theta(t)$, with $\theta(t)$ being the Heaviside step function. Initially, we compute the ground state without transverse magnetic field, which serves as our initial configuration, and subsequently we perform a real-time evolution.
The input and script files for this example can be found in the directory \verb|sample/07_timeevolution|. 

\begin{figure}[t] 
\begin{center} 
\includegraphics[width=0.8\columnwidth]{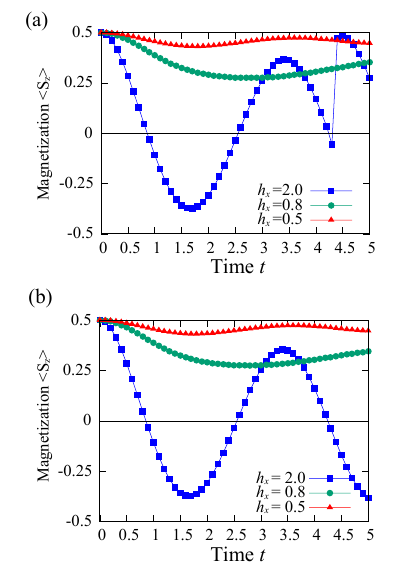}
\caption{Real-time evolution calculations of the Ising model with the transverse magnetic field $h_x = 0.5, 0.8, 2.0$; (a) the time dependence of the magnetization $\langle S_z \rangle$ at $D=3$ and (b) $D=10$.} 
\label{fig-TE} 
\end{center}
\end{figure}

\begin{figure}[t] 
\begin{center} 
\includegraphics[width=0.75\columnwidth]{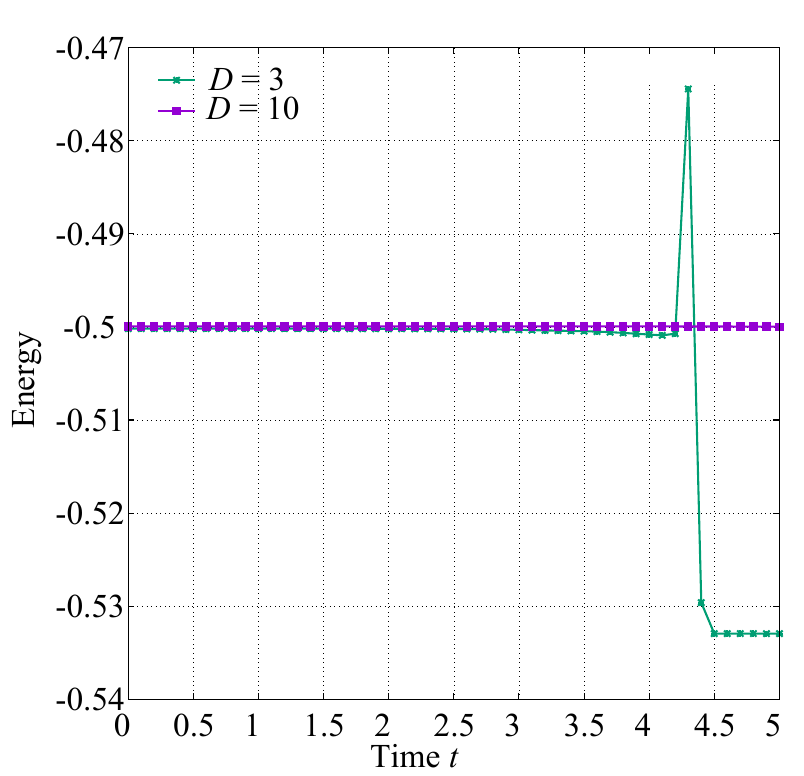}
\caption{Real-time evolution calculations of the energy on the Ising model with the transverse magnetic field $h_x = 2.0$ for $D=3$ and $10$.} 
\label{fig-TE-ene} 
\end{center}
\end{figure}

Figure~\ref{fig-TE}~(a) shows the numerical results obtained by \TeNeS when we take $D=3$. At $t=0$, the magnetization $\langle S_z \rangle$ is equal to $0.5$, since the initial state is the ferromagnetic state. 
After applying the transverse magnetic field $h_x$, the spins start to tilt so that they are magnetized in the $x$-axis direction, and the magnetization oscillates with increasing $t$. We can see that for $h_x = 2.0$, the magnetization jumps at $t=4.25$. This jump is also seen in the energy that should be conserved, as seen in Fig.~\ref{fig-TE-ene}. This is due to the increase of the quantum entanglement along the time evolution, i.e. the tensor network's capacity is insufficient to express the wave function. This problem can be resolved by increasing the \verb|virtual_dimension| $D$. 
In fact, this discontinuity disappears by setting $D = 10$ as seen from Figs.~\ref{fig-TE}~(b) and \ref{fig-TE-ene}. Users are advised to check the stability of the results against the change in the bond dimension for better reliability.

\subsection{Finite temperature}
\label{sec:finite_temp_example}

In this section, we present the calculation of the ferromagnetic Ising model on a square lattice with a transverse magnetic field, denoted by $h_x$, at finite temperatures.
The Hamiltonian is given as:
\begin{equation}
H = J^z \sum_{\langle i,j \rangle} {S}_i^{z} {S}_j^{z} - h_x \sum_i S_i^x.
\end{equation}
where $J^z = -1.0$. The step size $\tau$ for the imaginary time evolution operator is taken to be $\tau = 0.01$ for the first 50 steps, $\tau = 0.005$ for the following 200 steps, and $\tau = 0.05$ for the final 10 steps.
When the {\tt mode} parameter in the input file of \TeNeS is set to ``{\tt finite}'', we can perform tensor network calculations at finite temperatures. 
In this calculation, the bond dimension of CTM is set to $\chi=16$.

\begin{figure}[t]
  \begin{center}
    \includegraphics[width=1.0\columnwidth]{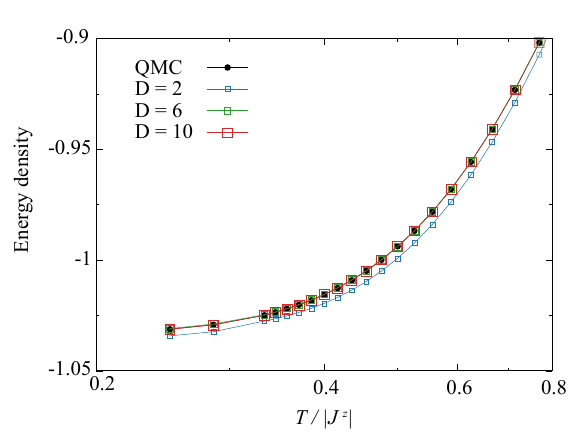}
    \caption{The energy density at $h_x = 2.0$ obtained by finite temperature calculations for $D=2,6$ and $10$ is shown as a function of the temperature. For comparison, the results obtained by the quantum Monte Carlo (QMC) method are also shown by filled circle symbols. The solid lines are for guiding eyes.}
  \label{fig:finite-temperature-ddep}
  \end{center}
\end{figure}

\begin{figure}[t]
  \begin{center}
    \includegraphics[width=1.0\columnwidth]{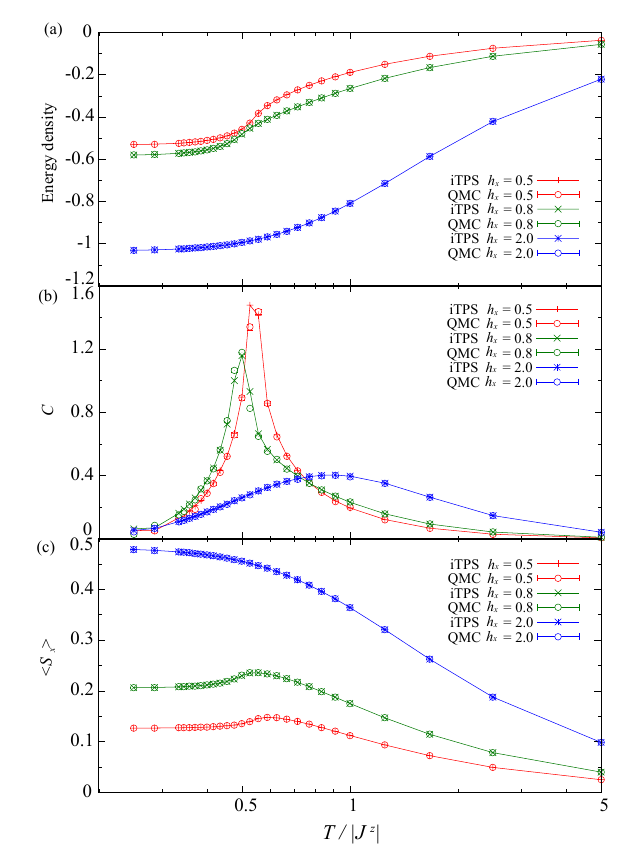}
    \caption{The results of finite temperature calculations for the Ising model with the transverse magnetic field: (a) the energy density, (b) the specific heat $C$, and (c) the magnetization $\langle S^x \rangle$. 
    These physical quantities are plotted as a function of the temperature $T/|J^z|$. For comparison, the results obtained by the quantum Monte Carlo (QMC) method are also shown by empty circle symbols. The solid lines are for guiding eyes.}
  \label{fig:finite-temperature}
  \end{center}
\end{figure}

Figure~\ref{fig:finite-temperature-ddep} shows the energy density at $h_x = 2.0$ as a function of the temperature for $D=2,6$ and $10$. The results obtained by the quantum Monte Carlo (QMC) method using {\tt DSQSS}~\cite{dsqss} are also shown by filled circle symbols for comparison.
As in real-time evolution, when the bond dimension is small, specifically $D=2$, the tensor network cannot accurately describe the wave function, resulting in deviations from the QMC values. For $D=6$ and $10$, the results are almost converged, and the discrepancy with QMC becomes negligible\footnote{For $D$ that can adequately represent the wave function, further increasing $D$ may cause numerical errors to accumulate and values to deviate. For example, in the case of the Ising model with $h_x =0$, a large deviation in energy values occurs when $D$ increases above $D=2$.}. Figures~\ref{fig:finite-temperature}(a-c) show the numerical results for (a) the energy density, (b) the specific heat, and (c) the magnetization in $x$-direction, as functions of the temperature $T/J^z$, when the bond dimension is taken as $D=10$.
The different symbols correspond to the different values of the transverse magnetic field $h_x$.
For comparison, the results obtained by the QMC method using {\tt DSQSS} are also shown by empty circle symbols.
The results obtained by these two methods coincide well with each other, indicating that the finite temperature calculations implemented in \TeNeS work well.

\subsection{Multi-sites observables}
\label{sec:multi_site_example}

For an example of calculations for multi-site observables, we consider the antiferromagnetic XXZ model on the triangular lattice under the magnetic field in the $z$-direction, whose Hamiltonian is given by
\begin{equation}
\mathcal{H} = \sum_{\braket{ij}} \left[ J^z S_i^z S_j^z + J^{xy} \left( S_i^x S_j^x + S_i^y S_j^y \right) \right] - h\sum_i S_i^z.
\end{equation}
For $J^{xy} > J^{z} \ge 0$, there appears a non-coplanar (umbrella) phase under a weak magnetic field, in which three spins on each small triangle are not on the same plane~\cite{Yamamoto2014-tc, Sellmann2015-qd}.
This phase can be characterized by the scalar chirality density,
\begin{equation}
\chi = \frac{1}{N}\left(\sum_{\braket{ijk} \in \set{\bigtriangleup}} \chi_{ijk} - \sum_{\braket{ijk} \in\set{ \bigtriangledown}} \chi_{ijk}\right),
\end{equation}
where $\chi_{ijk}$ is the scalar chirality of three spins defined in Eq.~(\ref{eq:chirality})
and $\braket{ijk}\in \set{\bigtriangleup(\bigtriangledown)}$ means that $i,j$, and $k$ sites form an upward (downward) triangle in the counter-clockwise order. To calculate the scalar chirality density in iTPS, we consider the sum of all upward and downward triangles within a unit cell. 

\begin{figure}[t]
\includegraphics[width=.9\linewidth]{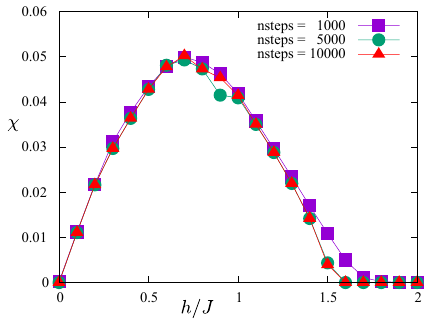}
\caption{Scalar chirality density of the antiferromagnetic XY model on the triangular lattice under the magnetic field along the $z$-axis. The solid lines are for guiding eyes.}
\label{fig:chirality}
\end{figure}

Figure~\ref{fig:chirality} shows the magnetic field dependence of the scalar chirality density $\chi(h/J)$ of the XY model ($J^z = 0, J^{xy} = 1$) obtained by the simple update method with the bond dimension $D=3$ and iTPS is contracted by using the mean-field environment.
The initial configuration, the 120-degree ordered state, breaks the chiral (mirror-image) symmetry while making the scalar chirality zero.
When the magnetization field $h/J$ is weak, the scalar chirality is well converged to finite values. Note that for $h/J > 1.3$, we observe a slow convergence with respect to the imaginary time evolution steps, indicating some difficulties in the high-field simulation. Indeed, when we increase the bond dimension to $D=4$, we observe an unexpected increase in the energy and a peculiar singularity in the scalar chirality around $h/J = 1.4$. One of the reasons for such unphysical behavior might be the simple update, which only considers the mean-filed environment, combined with the simplified treatment of diagonal interaction when we map the triangular lattice into the square lattice \cite{TeNeSv1, TeNeS_webpage}.

\section{Conclusion}
\label{sec:conclusion}

In this paper, we introduced the new features implemented in version 2 of \TeNeS. \TeNeS is applicable to a variety of quantum spin and boson models on arbitrary infinite two-dimensional lattices. With the new version, ground state calculations, real-time evolutions, and finite temperature simulations can be performed.

The real-time evolution is a simple extension of the imaginary-time evolution often used in the optimization for the ground state. Due to the exponential growth of the entanglement along the real-time evolution, even if the simulation is started from a low-entangled state, the accurate approximation by the iTPS/iPEPS representation is limited to relatively short time intervals. However, we can extract meaningful information from short-time iTPS/iPEPS simulations in some cases \cite{HubigBKGC2020,KanekoD2022}. The real-time simulation mode in \TeNeS-v2 may support future studies in such directions.

The finite temperature simulation is another important direction for tensor-network-based calculations. Recent tensor-network simulations of the Shastry--Sutherland model have shown that we can obtain accurate specific heat even in the vicinity of the quantum critical point, and these results can be used to understand experiments for $\mathrm{SrCu_2(BO_3)_2}$ \cite{Jimenez2021}. Using the new features in \TeNeS-v2, one can conduct similar finite temperature simulations. 

Since \TeNeS-v2 supports real-time evolutions and finite temperature simulations with only a few modifications of the input file from the ground state calculation, users can easily apply these new features to their own models. For users who want to easily experience these features, \TeNeS is pre-installed in the MateriApps LIVE! virtual environment\cite{MOTOYAMA2022101210}, making it easy to get started.
We hope that \TeNeS will stimulate new researches in quantum many-body systems on two-dimensional lattices.

\section*{Acknowledgments}

\TeNeS was developed with the support of the ``Project for advancement of software usability in materials science'' (PASUMS) in fiscal year 2019 and 2023 by The Institute for Solid State Physics, The University of Tokyo. This work is supported by Japan Society for the Promotion of Science KAKENHI, Grant Nos. 20H00122, 22H01179, 22K18682, 23K22450, 23H03818, 23H01092, 23K25789 and by the Center of Innovations for Sustainable Quantum AI (JST Grant Number JPMJPF2221).
T. O. acknowledges the support from the Endowed Project for Quantum Software Research and Education, The University of Tokyo (https://qsw.phys.s.u-tokyo.ac.jp/).
The computations in this work were done using the facilities of the Supercomputer Center, The Institute for Solid State Physics, The University of Tokyo.


\bibliographystyle{elsarticle-num}

\end{document}